\documentclass[a4paper,12pt,useAMS,usenatbib]{mn2e}

\usepackage{graphicx}
\newcommand{\aap}{A\&A}
\newcommand{\apj}{ApJ}
\newcommand{\apjl}{ApJ}
\newcommand{\apjs}{ApJS}
\newcommand{\aj}{AJ}
\newcommand{\pasj}{PASJ}

\newcommand{\mnras}{MNRAS}

\newcommand{\araa}{ARA\&A}
\newcommand{\nat}{Nature}
\newcommand{\aaps}{A\&AS}

\title[Cluster-Cluster Lensing: A383]{Cluster-Cluster Lensing and the Case of Abell 383}

\author[Zitrin et al.]{Adi Zitrin$^{1}$\thanks{E-mail:
adiz@wise.tau.ac.il}, Yoel Rephaeli$^{1}$, Sharon Sadeh$^{1}$, Elinor
Medezinski$^{2}$, Keiichi Umetsu$^{3}$, \and Jack Sayers$^{4}$, Mario Nonino$^{5}$, Andrea Morandi$^{1}$, Alberto Molino$^{6}$, Nicole
Czakon$^{4}$, \and Sunil R. Golwala$^{4}$\\\\\\
$^{1}$School of Physics and Astronomy, the Raymond and Beverly Sackler
Faculty of Exact Sciences, Tel Aviv University,\\ Tel Aviv 69978, Israel\\
$^{2}$Department of Physics and Astronomy, The Johns Hopkins University, 3400 North Charles Street, Baltimore, MD 21218\\
$^{3}$Institute of Astronomy and Astrophysics, Academia Sinica, P.~O.
Box 23-141, Taipei 10617, Taiwan\\
$^{4}$Division of Physics, Mathematics, and Astronomy, California Institute of Technology, Pasadena, CA 91125\\
$^{5}$INAF-Osservatorio Astronomico di Trieste, Via Tiepolo 11, I-34131 Trieste, Italy\\
$^{6}$Instituto de Astrof\'isica de Andaluc\'ia (CSIC), C/Camino Bajo de Hu\'etor, 24, Granada, 18008, Spain\\
}

\begin{document}


\pagerange{\pageref{firstpage}--\pageref{lastpage}} \pubyear{2011}

\maketitle

\label{firstpage}

\begin{abstract}

Extensive surveys of galaxy clusters motivate us to assess the likelihood of cluster-cluster lensing (CCL), namely, gravitational-lensing of a background cluster by a foreground cluster. We briefly describe the characteristics of CCLs in optical, X-ray and SZ measurements, and calculate their predicted numbers for $\Lambda$CDM parameters and a viable range of cluster mass functions and their uncertainties. The predicted number of CCLs in the strong-lensing regime varies from several ($<10$) to as high as a few dozen, depending mainly on whether lensing triaxiality bias is accounted for, through the c-M relation. A much larger number
is predicted when taking into account also CCL in the weak-lensing regime. In addition
to few previously suggested CCLs, we report a detection of a possible CCL in
A383, where background candidate high-$z$ structures are magnified, as seen in deep
Subaru observations.
\end{abstract}


\section{Introduction}

The mass density in the central regions of galaxy clusters typically exceeds
the critical value required for lensing, generating multiple-images of
background objects. This phenomenon is known as strong-lensing (SL) and the
background sources are usually very distant field galaxies, lensed into
magnified and often multiple arcs on the lens plane. Recent analyses have
shown that many sets of multiply-lensed images can be uncovered with
high-quality space imaging measurements and
improved modelling techniques \citep[e.g.,][]{Broadhurst2005a,Liesenborgs2007,
Limousin2008,Newman2009,Zitrin2009b,Coe2010,Deb2010Abell901,Richard2010locuss20,
Merten2011}.

With more precise knowledge of the global and large-scale parameters, and
extensive ongoing surveys of galaxy clusters in several spectral regions,
the possibility of a foreground cluster lensing a background cluster is of
practical interest.

An initial estimate of the possibility of observing cluster-cluster lensing (CCL) was made by \cite{Cooray1999CCL}, who predicted
that a few dozen CCLs may be observed over the full sky. Soon thereafter, two such lenses
were discovered. \cite{BlakesleeCCL2001} and \cite{Blakeslee2001Hercules} found that the
nearby supercluster A2152 ($z =0.043$) is actually a chance alignment of two clusters: A2152, and a more massive
background cluster at $z =0.134$ (which was then designated A2152-B). The centres of
these two clusters are separated by 2.4$\arcmin$, and some background cluster galaxies of
the more distant cluster seem magnified and distorted in the image-plane of A2152.
\cite{Athreya2002MS1008} have shown that an excess of distant galaxies in the
South-West area of MS 1008-1224 is most likely also a weaker lensing effect
of a background cluster near the line of sight. There seem to be no other explicit cases of CCLs reported to date.

\cite{BertinLombardi2001CCL} have also investigated the properties of a ``double lens'' configuration, and mainly its effect on WL analyses. In this context, the effect of interest is the lensing of a background source by two (at least partially) aligned lenses, where several such configurations were suggested or theoretically discussed before \cite[e.g.,][]{Crawford1986doubleLens,SeitzSchneider1994,Molinari1996doubleLens,Wang1997doubleLens,Gavazzi2008doubleLens}.

Major advances in the capability of detecting weak low-brightness emission in
the optical and X-ray regions, and more recently also in SZ mapping of many
clusters, together with increased precision in the values of the cosmological
parameters, make the (strong) CCL phenomenon of practical interest as a probe of
cluster properties. Additionally, the statistics of CCLs enhances the use of
clusters as probes of the evolution of the large scale structure (LSS).
The manifestation of SL in different bands of the electromagnetic spectrum
secures the identification of CCLs. Contrasting the results from a search of
CCLs with theoretical predictions may yield important new insight, especially
on the late evolution of the LSS.

We briefly describe the possible observational signatures of CCLs in the optical,
X-ray and SZ, and carry out a detailed calculation of the expected numbers
of CCLs in several cosmological models using current values of the global
and cluster parameters. Our updated treatment here (for previous calculation
see \citealt{Cooray1999CCL}) yields a wide range of values for the predicted
numbers of CCLs, reflecting modelling and observational uncertainties.
Additionally, we report a (possible) discovery of another moderately-lensed
(magnified by $14\pm3~ \%$; see \S \ref{383}), high-$z$ background cluster at
$z\sim0.9^{+0.2}_{-0.1}$ behind A383 ($z=0.19$), $\sim2.2\arcmin$ from its centre, as seen in Figure \ref{CC383}.

The paper is organised as follows: In \S 2 we discuss the observational
properties of CCL. In \S 3 we present our calculation of the probability
for CCLs and their predicted numbers. The possible detection of a CCL in
the field of A383 is discussed in \S 4. Our main results are summarized
in \S 5.

\section{Observational Properties Of CCL\lowercase{s}}

Lensing of a background cluster results in magnified optical images of the background
cluster galaxies, and in hitherto undetected signatures in the X-ray and microwave
regions.
In the image plane, a clear local overdensity of magnified, distorted, and
stretched optical images would generally be expected when the galaxies of a
background cluster are lensed.
The higher redshift of the background cluster should result in
images fainter by the luminosity distance ratio (relative to the lensing
cluster), but boosted by the magnification effect which though preserves
surface brightness, will magnify the total flux (due to the increased area occupied
by each source in the image-plane). Also, when having multi-band imaging, the higher
redshift of the background cluster will cause the background galaxies to look redder
relative to the lens red-sequence galaxies (see Figure \ref{CC383}), though this
effect might not be prominent in a simple RGB colour-composite image when redshift
differences are relatively small (and further weakened by the Butcher-Oemler effect;
e.g., \citealt{ButcherOemler1978}), and in such a case are more likely to be revealed
by producing photometric catalogues which may exhibit a different, secondary
red-sequence corresponding to the background cluster.

The steep dependence of the X-ray surface brightness on redshift would
generally mean that, even though magnified by the foreground cluster, the background
cluster will at best look as a faint part of the foreground cluster emission. If the
background cluster lies further away from the line of sight, there might be a
traceable signature, as the magnified background flux will be seen far enough from
the foreground cluster centre, where the foreground flux is lower and thus might
enable a clear detection, but only if the background cluster is sufficiently luminous.
We note, however, that in the strong regime this may resemble X-ray images of a
substructure, merger, or related shocks, and thus without additional information,
even if such a signal is detected it could well be miss-interpreted.
In addition (as was noted previously by \citealt{Cooray1999CCL}), X-ray spectra can
be used to determine the background cluster redshift, particularly by the measurement
of the relatively strong Fe lines. Current measurement capabilities (e.g, with
\emph{Chandra}) enable determining the cluster redshift up to
$z\sim1$.
Still, it might not be feasible to detect a CCL based solely on X-ray imaging
measurements.

The SZ effect is the change in the CMB intensity due to
Compton scattering of CMB photons as they traverse intracluster gas \citep[e.g.,][]{Rephaeli1995,Carlstrom2002SZreview}.
The result is a redshift-independent distortion of the CMB spectrum,
whose thermal component constitutes a decrement of CMB spectrum below
$\simeq 218$~GHz, and an increment of the spectrum above $\simeq 218$~GHz.

The SZ effect is measured with respect to the unscattered CMB at the location of the
cluster, so lensing affects the SZ signal similarly at all observed frequencies.
The total observed SZ signal is therefore a sum of the intrinsic SZ signal from the
lensed cluster, which is then magnified by the foreground cluster, plus the SZ signal
due to scattering in
the foreground cluster.

The fact that the SZ effect is independent of redshift may help in making the
identification of a CCL more feasible in SZ surveys than in X-ray surveys. However, it
may still be hard to disentangle the signals of the foreground and background
structures, as the SZ signal usually stretches out to large (projected) distances and may thus cover-up lensed features. The net result from such a trade-off will depend on the lens and source redshifts, and on the projected distance of the lensed feature from the line-of-sight of the lensing cluster, so that generally, structures closer to the line-of-sight will be more strongly lensed and magnified, but the relative flux from the foreground cluster will also be higher. As in the X-ray, also here strong CCL may resemble images of merger, or related shocks. Obviously, a CCL identification can be more secure if lensing features are revealed in both X-ray and SZ measurements, or if clearly detected in optical imaging measurements.
With photometric and/or spectroscopic data, such a detection could yield precise
information on both clusters.

In \S 4 we elaborate further on the lensing signatures in the optical, X-ray
and SZ regions, as manifested in A383, which
seems to be moderately lensing
a background cluster.

\section{Predicted Numbers of Cluster-Cluster Lenses}

In order to assess the probability for CCLs we integrate the mass function over two
cluster populations, namely those of the lenses and sources. We do so for both
the \cite{PressSchechter1974} mass function (hereafter PS), and that of
\cite[hereafter ST]{ShethTormen1999}. For each cluster of the lens population we
integrate the mass function of sources lying behind the lens, and included within
the Einstein radius, as (properly) determined by the source and lens redshifts.
The volume element over which the source integration is carried out is computed in
terms of the solid angle defined by the Einstein ring in the plane of the lens,
projected onto the source redshift, by means of the ratio of the squared lens-source
angular diameter distances. Integrating over the lens population then yields the
desired number of CCLs.

Note that we include in this estimate only sources that are fully enclosed
within the respective Einstein radius, ignoring a partial alignment of
the source within this radius. In this regard our estimate constitutes
a lower limit on the number of CCLs.

\subsection{Method}

The equation governing the relation between the concentration parameter and
the Einstein radius assuming an NFW profile \citep{BroadhurstBarkana2008} is,
\begin{equation}
\left(\frac{4R_v\rho_c^z\Delta_c}{3\Sigma_{cr}}\right)
\frac{c_v^2}{\ln{(1+c_v)}-c_v/(1+c_v)}\frac{g(x)}{x^2}=1,
\label{eq:gx}
\end{equation}
where $\Delta_c$ and $\rho_c^z$ are the overdensity at virialisation and
critical density at redshift $z$, respectively,
\begin{equation}
\Sigma_{cr}=\frac{c^2}{4\pi G}\frac{D_{OS}}{D_{OL}D_{LS}}
\end{equation}
is the critical surface density,
$D_{OS}$, $D_{OL}$, and $D_{LS}$ are the observer-source, observer-lens, and
lens-source distances, respectively, and
\begin{equation}
R_v=\frac{1.69}{1+z}\left[\frac{M}{M_{15}}
\frac{18\pi^2}{\Omega_m\Delta_c(\Omega_m,z)}\right]^{1/3} Mpc\cdot h^{-1},
\end{equation}
is the virial radius. Also, $x\equiv\frac{R_E c_v}{R_v}$,
where $R_E$ is the Einstein radius, and
\begin{eqnarray}
g(x)=\ln{\frac{x}{2}}+
\left\{\begin{array}{ll}
1, &\mbox{$x=1$} \\
\frac{2}{\sqrt{x^2-1}}\tan^{-1}\sqrt{\frac{x-1}{x+1}}, &\mbox{$x>1$} \\
\frac{2}{\sqrt{1-x^2}}\tanh^{-1}\sqrt{\frac{1-x}{1+x}}, &\mbox{$x<1$}
\end{array}\right..
\end{eqnarray}

The concentration parameter $c_v$ scales with mass and redshift according to
the following relation:
\begin{equation}
c_v= A (M/M_{*})^B (1+z)^{C},
\label{cm}
\end{equation}
where the parameters A, B, C and $M_{*}$ are taken from various c-M relations
as we discuss below.

The solution of Eq.~(\ref{eq:gx}) provides the Einstein radius as a function
of $c_v$, from which the angular Einstein radius $\theta_E=R_E/D_A(z_{l})$, the ratio
between the (physical) Einstein
radius and the angular diameter distance to the lens, can be readily determined.

Having solved for the angular Einstein radius at the lens, we can now
estimate the number of source clusters that would undergo lensing, i.e. the
ones lying behind the lens and included within the angular area subtended
by the Einstein radius. For this purpose we set the lens mass and redshift,
and integrate the mass function over the relevant mass range, a volume element
defined by the source-cluster redshift, $z_{L}< z_{S}<\infty$, and the angular
diameter element specified by the Einstein radius of the lens-source system,
projected onto the source redshift by means of the squared lens-source angular
diameter distance ratio, $(d_{A_L}/d_{A_S})^2$ (Fig. 1). This result provides the
number of lensed sources behind a lens lying at redshift $z_L$, and having a mass
$m_L$. The total number of CCL occurrences is likewise
estimated by integrating the mass function over the mass-redshift space of the
lens:

\begin{tiny}
\begin{eqnarray}
N_{cc}={}&\int_{m_L}\int_{V_L}\int_{m_s}\int_{V_S}
\,n(m_L,z_L)n(m_S,z_S)dm_LdV_Ldm_SdV_S\\
={}&4\pi\int_{m_L}\int_{z_L}n(m_L,z_L)r_L^2dm_L\frac{dr_L}{dz_L}dz_L...\nonumber\\
...{}&\int_{m_S}\int_{z_S}\int_{\Omega_{\theta_E}}n(m_S,z_S)dm_Sdz_S
d\Omega_{\theta_E}
\nonumber.
\end{eqnarray}
\end{tiny}

\begin{figure}
\centering
\includegraphics[width=80mm, clip]{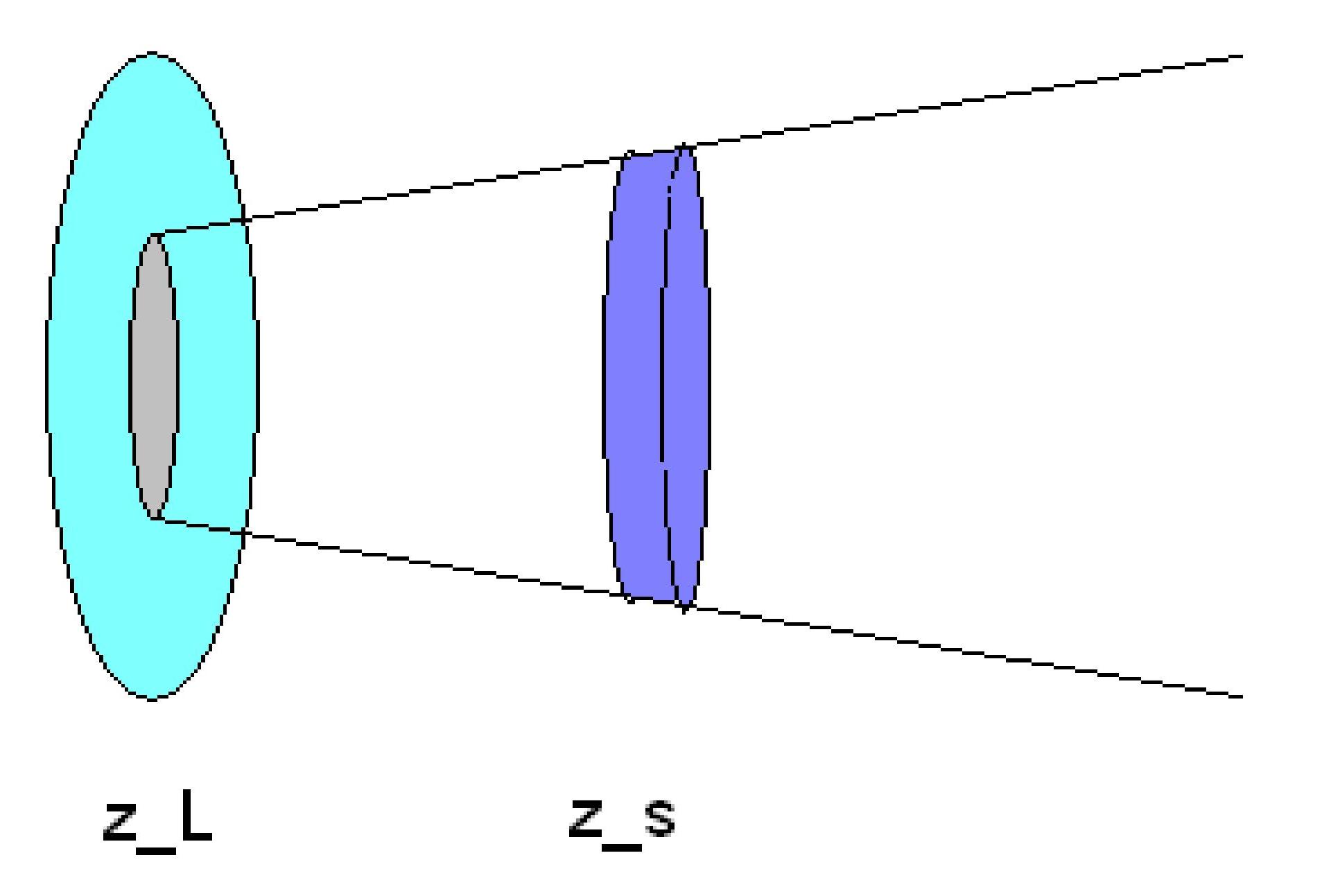}
\caption{The lens-source configuration. The grey area represents the section
of the lens plane which is included within the Einstein radius, as calculated
from the lens and source redshifts. Integration of the source population is
performed along the cone extending to the right of the lens, with an angular
cross section corresponding to the projected Einstein radius at the source
redshift, $z_S$. The corresponding source-plane volume element is denoted in
purple. Note that by virtue of the low angular scales involved, one can
safely use the flat sky approximation to calculate the volume element at
the source redshift.}
\end{figure}

\subsection{Results}

As is obvious, results depend significantly on the mass function, and
quite strongly on the c-M relation. This relation is only roughly estimated
from numerical simulations of clusters; therefore, it introduces a large
uncertainty in the predicted numbers of CCLs. We have used the PS and ST mass
functions and various c-M relations, each specified in terms of a different set
of A,B,C and $M_{*}$ parameters (in eq. \ref{cm}). We first use the notation and
parameters given in \cite{KomatsuSeljak2002}, based on the work of
\cite{Seljak2000} and \cite{Bullock2001}. In this notation
$M_{*}=5.2\times 10^{12}M_{\odot}$
(as calculated by us according to WMAP7 parameters) is the solution to
$\sigma(M)=\delta_{c}$, where $\sigma(M)$ is the present day rms mass
fluctuations, and $\delta_{c}$ the threshold overdensity for spherical collapse
at $z=0$, with $A=10$, $B=-0.2$, and $C=-1$.

In a $\Lambda$CDM cosmological model
with $(\Omega_m,\Omega_{\Lambda},n,h,\sigma_8)=(0.266,0.734,0.963,0.71,0.801)$
as taken from WMAP7 results, the above parameters for the c-M relation,
and clusters in the mass interval,
$1\times 10^{13}M_{\odot}$ - $1\times 10^{16}M_{\odot}$
our calculations yield $\sim0.03$ CCLs with a PS mass function, and $\sim0.1$
CCLs with a ST mass function. When taking into account also background groups of
galaxies down to $5\times 10^{12}M_{\odot}$, we obtain $\sim0.1$ CCLs with a PS
mass function, and $\sim0.5$ CCLs with the ST mass function.

We repeated the calculation with the c-M relation given by \cite{Duffy2008} for
their full sample, in which $M_{*}=2\times 10^{12} h^{-1} M{\odot}$, $A=7.85$,
$B=-0.081$, and $C=-0.71$. With these values, for a $\Lambda$CDM
model with $(\Omega_m,\Omega_{\Lambda},n,h,\sigma_8)=(0.258,0.742,0.963,0.719,0.796)$
taken from WMAP5 results (as used in Duffy et al. 2008), and mass limits
of $1\times 10^{13}M_{\odot}$ - $1\times 10^{16}M_{\odot}$, our calculations yield
$\sim1$ CCLs with a PS mass function, and $\sim2$ CCLs with a ST mass function.
When taking into account also background groups of galaxies down to
$5\times 10^{12}M_{\odot}$, our calculations yield $\sim3$ CCLs with a PS mass function,
and $\sim8$ with a ST mass function. We note that the c-M relation presented in
\cite{Bullock2001} yields similar results.

In order to take into account the lensing projection bias of
triaxial cluster morphology,
the calculation was repeated with the c-M relation of \cite{Hennawi2007}, in which
$M_{*}=1.3\times 10^{13} h^{-1} M{\odot}$, $A=12.3$, $B=-0.13$, and $C=-1$.

For these values, in $\Lambda$CDM model with $(\Omega_m,\Omega_{\Lambda},n,h,
\sigma_8)=(0.258,0.742,0.963,0.719,0.796)$ taken from WMAP5 results, and mass limits
of $1\times 10^{13}M_{\odot}$ - $1\times 10^{16}M_{\odot}$, our calculations yield
$\sim9$ CCLs with a PS mass function, and $\sim17$ CCLs with a ST mass function.
When taking into account also background groups of galaxies down to
$5\times 10^{12}M_{\odot}$, we predict $\sim38$ CCLs with a PS mass function, and $\sim68$ CCLs with a ST mass function.
Thus, taking into account the lensing projection bias boosts CCL numbers by about an
order of magnitude.

We compare these results to the observed c-M relation from a small sample of 10 clusters derived by \cite{Oguri2009}, in which $M_{*}=1\times 10^{15}
M{\odot}$, $A=12.4$, $B=-0.081$, and $C=-1$. This relation yields concentrations
higher than predicted by $\Lambda$CDM simulations, and even higher than those
derived observationally by previous work (e.g., \citealt{
ComerfordNatarajan2007CMrelation}), and are likely to be extreme results that are perhaps less relevant for our purposes.

Assuming these values, in $\Lambda$CDM with $(\Omega_m,\Omega_{\Lambda},n,h,\sigma_8)
=(0.258,0.742,0.963,0.719,0.796)$ taken from WMAP5 results, and mass limits of
$1\times 10^{13}M_{\odot}$ - $1\times 10^{16}M_{\odot}$, our calculations yield
$\sim165$ CCLs with a PS mass function, and $\sim260$ CCLs with a ST mass function.
When taking into account also background groups of
galaxies down to $5\times 10^{12}M_{\odot}$, our calculations yield $\sim660$
CCLs with a PS mass function, and $\sim940$
CCLs with a ST mass function. Thus, according to the observed relation (which is
known to produce higher concentration than $\Lambda$CDM simulations) the predicted total numbers of CCLs are quite large.

It should be noted that taking into account the weak lensing regime, in which the
background cluster does not have to be within the Einstein radius as projected
onto the source plane, but can be further out up to several Einstein radii, the
likelihood of a CCL increases significantly. Specifically, still assuming the flat
sky approximation, the likelihood of a CCL increases simply as the square of the
ratio of the projected distance of the background cluster and the Einstein radius
of the foreground cluster.

Moreover, extending the calculations to the 1$\sigma$ ranges of the $\Lambda$CDM
parameters broadens the ranges of our predicted numbers of CCLs by up to a factor of
$\sim 2$. Finally, the use of other mass functions - such
as \cite{Jenkins2001MF} or \cite{Tinker2008MF}
- can introduce another $\sim$20\% variation.

\begin{figure*}
 \begin{center}
   \includegraphics[width=160mm]{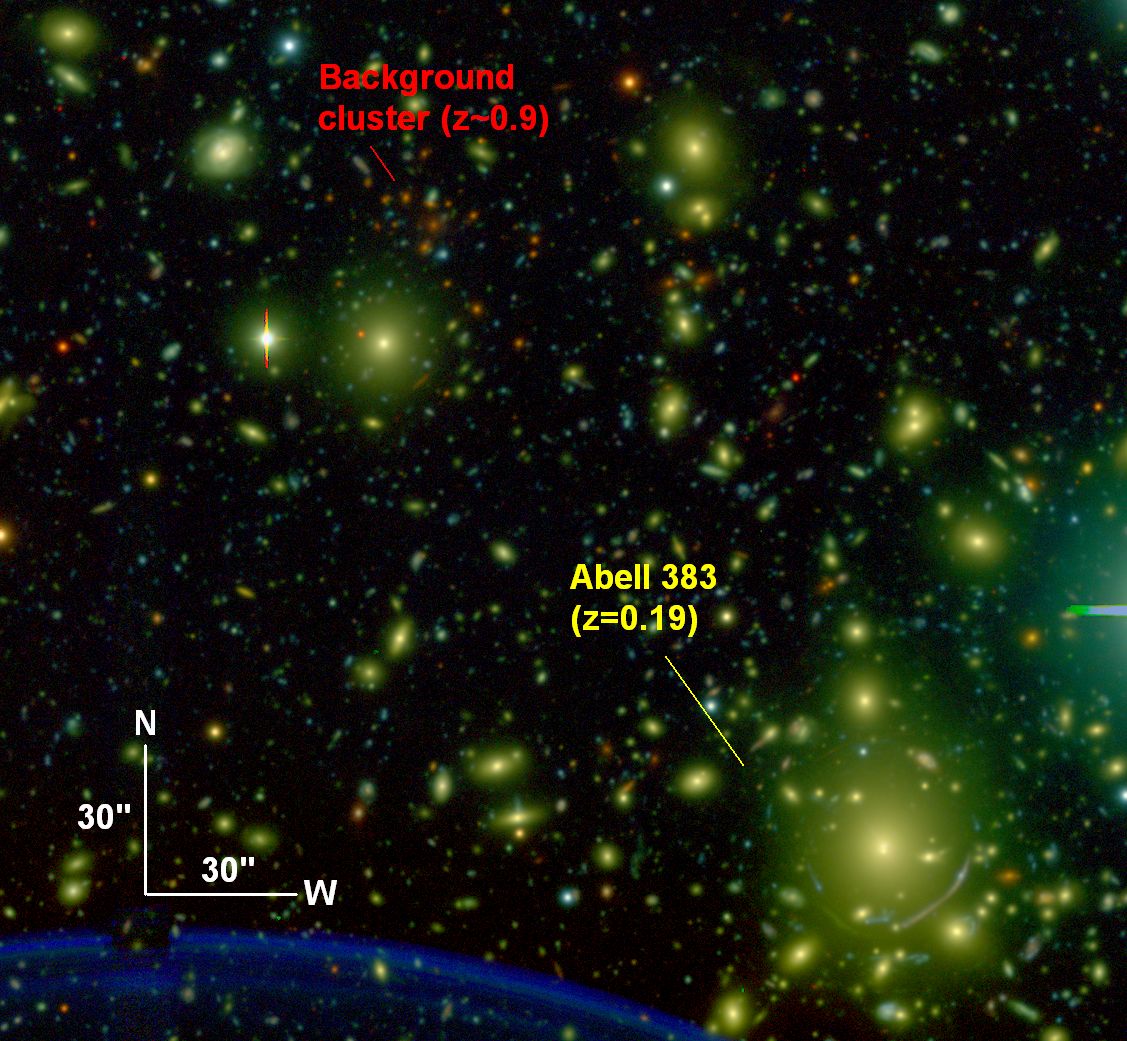}
 \end{center}
\caption{The central field of A383 with the high-$z$ background cluster marked on
the image. Note the much redder colours of the galaxies in the background cluster
with respect to A383. According to our WL analysis, A383 magnifies the background
cluster by $14\pm3 \%$ (see \S4 for more details).}
\label{CC383}
\end{figure*}

\section{A383: A New CCL?}\label{383}

In the course of this work we inspected (among other clusters) deep
archival multiband images of A383 obtained with the SuprimeCam on the
Subaru telescope \citep{Miyazaki2002SuprimeCam} in 2002,2005,2007,2008 and new data
collected in 2010 dedicated to the CLASH sample (\citealt{PostmanCLASHoverview}; see
also below), with total integration times of at least $\sim1$hr and up to $\sim4$hr, for
each of the B,V,Rc,Ic,i',z' bands. Our analysis indicates that this is possibly a new
example of a CCL, with the lensed system being either a background cluster or a group of galaxies.
Standard image reduction was performed with \emph{mscred} task in IRAF
\footnote{\cite{Valdes1998MSCRED}. IRAF is distributed by the National Optical
Astronomy Observatories, which are operated by the Association of Universities for
Research in Astronomy, Inc., under cooperative agreement with the National
Science Foundation.}, while coadded images were created following
\cite{Nonino2009ReductionSubaru}. Zero-points were estimated from standard stars observations.

In these wide-field ($\approx 30\arcmin \times 30\arcmin$) Subaru images of the
field of A383, the multiwavelength coverage uncovers several higher-$z$ structures. Among these,
two large structures are seen; one is located $\sim5.2\arcmin$ East and
$\sim2.2\arcmin$ North of A383, around RA=02:48:24.96 DEC=-03:29:31.8, and the
second $\sim13\arcmin$ East and $\sim2.8\arcmin$ North of A383, around RA=02:48:56.03
DEC=-03:29:06.6, and extends Northwards towards a third (possibly different)
substructure of similar colours (and redshift), but these structures are too far
from the centre of A383 to be relevant for this work. Our BPZ \citep{Benitez2000,
Benitez2004, Coe2006} photometric catalogue, based on the 6 Subaru imaging bands
mentioned above, suggests redshifts of $z\sim0.3$ and $z\sim0.7$ for these
structures, respectively, which further reduces their expected magnification by
A383 to $\simeq1$.

The more interesting case we consider here
is a clear redder sequence of galaxies, $\sim2.5\arcmin$ North-East of the centre
of A383, around RA=02:48:09.57 DEC=-03:29:41.6 (see Figure \ref{CC383}). The background structure is clearly seen as a red overdensity of galaxies
very faint in the B-band (see Figure \ref{CC383} for an RGB colour image), implying a higher redshift than A383, further
confirmed by a photometric redshift of $z\sim0.9^{+0.2}_{-0.1}$ for all 40
member galaxies. The average photometric redshift for these galaxies is $z=0.96$,
though we adopt the photo-$z$ of the brightest member, $z\simeq0.9$, as the
redshift of this structure.

\begin{figure}
 \begin{center}
   \includegraphics[width=80mm]{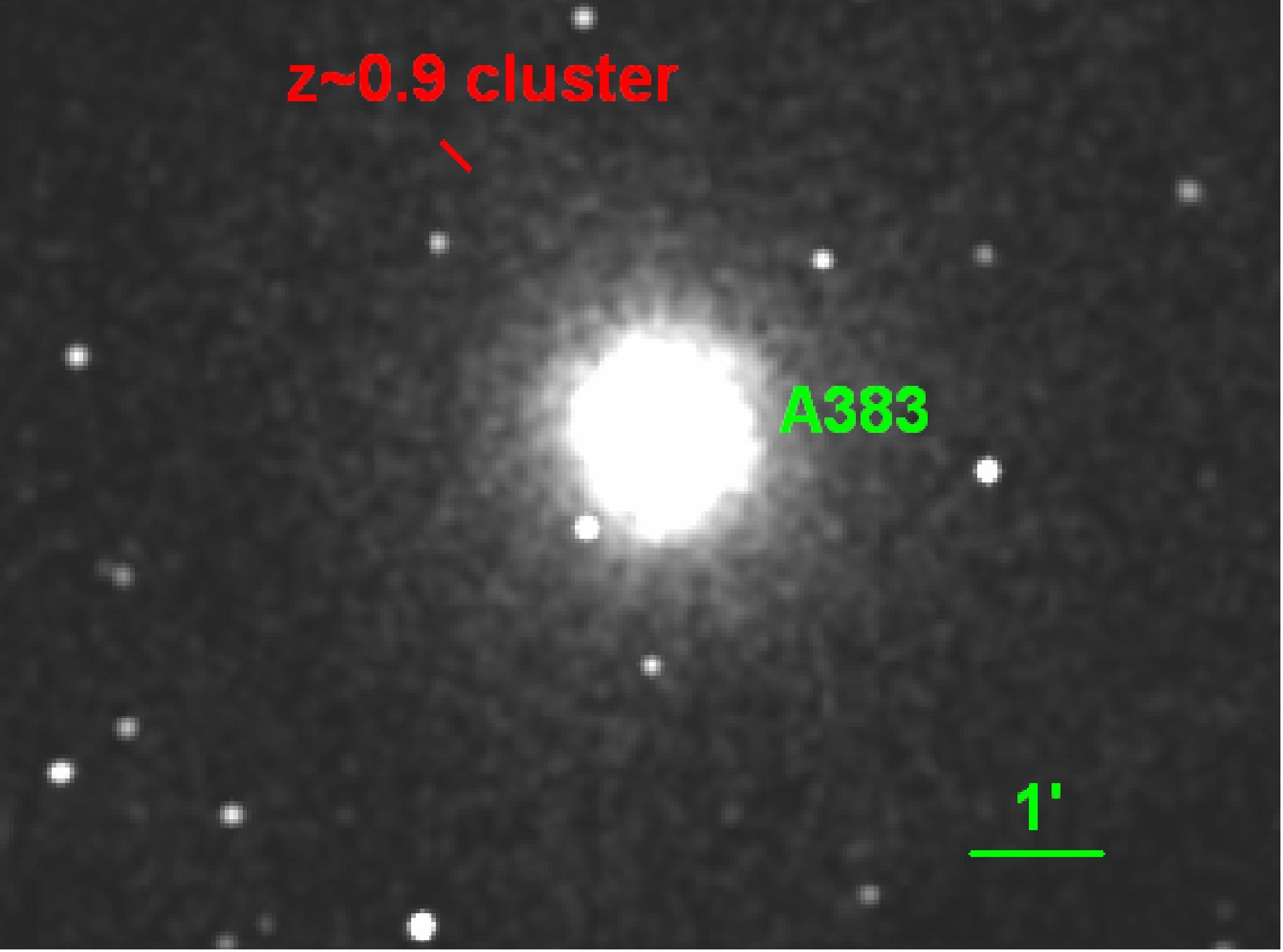}
 \end{center}
\caption{Smoothed X-ray image of
A383. The
location of the z$\sim0.9$ background
cluster is indicated by the red line.
No noticeable excess is seen in that location, as can be expected (see \S 2).}
\label{xray383}
\end{figure}

We have downloaded $Chandra$ X-ray images and obtained Bolocam 140~GHz SZ images of
A383 to examine whether this background cluster is seen also in these spectral
regions. As may be expected (see \S 2), no excess X-ray brightness is seen at the location of the cluster, possibly due to a low mass
(and temperature) cluster or group. Neither is the lensed cluster detected in the
processed SZ map. The optical, SZ, and X-ray signals of the lensed cluster,
are used below to place limits on its mass, along with a WL analysis mass-estimation.

A383 is one of 25 clusters covered in the CLASH multi-cycle treasury program. The
CLASH program has been awarded 524 orbits of HST time to conduct a multi-cycle
program that will couple the gravitational-lensing power of 25 massive intermediate
redshift clusters with HST's newly enhanced panchromatic imaging capabilities
(WFC3 and the restored ACS), in order to test structure formation models
with unprecedented precision. More details can be found in \cite{PostmanCLASHoverview}.

To further base the results of this work, and in the framework of the CLASH
collaboration, we have constructed strong and weak lensing (WL)
models for this field (respectively, \citealt{Zitrin2011c}, Umetsu et al. in
preparation). Interestingly, the different background structures are seen in the
wide-field WL analysis, which we use to derive a magnification of $14\pm3~ \%$ at the location of
the lensed cluster discussed here. We note that this cluster is $\sim10$ Einstein radii
away from the centre of A383, therefore the SL model cannot be used for the
magnification estimate, but only to check consistency with our WL analysis of A383 in
the region of overlap (around 0.7-1 arcminute in radius; see Figure 5 in
\citealt{Zitrin2011c}).

\begin{figure}
 \begin{center}
   \includegraphics[width=80mm]{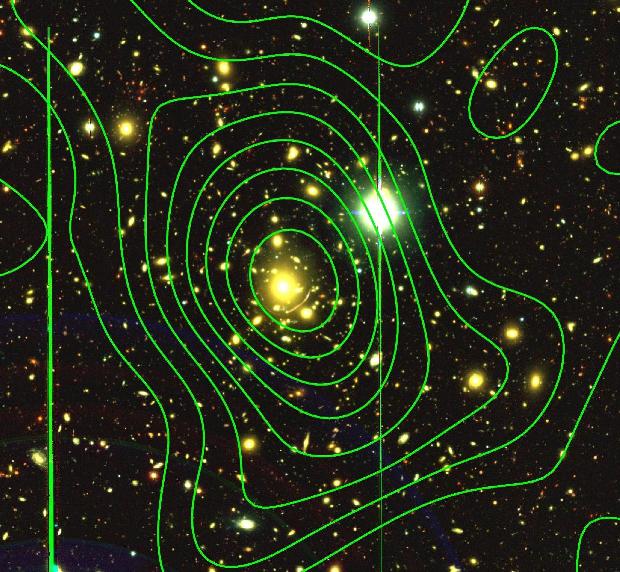}
 \end{center}
\caption{Contours of the dimensionless surface mass density $\kappa$ smoothed
 with a Gaussian of FWHM $1.7\arcmin$, superposed on the $BRz'$ pseudocolour
 image of A383.
The lowest contour and the contour interval are $\Delta\kappa=0.03$,
 corresponding to the $1\sigma$ reconstruction error.
As can be seen, the WL analysis is sufficiently sensitive to detect
a clear extension around the location
of the lensed background cluster detected in the optical deep RGB colour in
Figure \ref{CC383}.
}
\label{CC383WL}
\end{figure}

It should be noted, that these higher-$z$ structures are deduced
in both 1D and 2D WL analyses, even though different approaches are employed in these.
In the 1D analysis (e.g., \citealt{Umetsu2011}), such background structures are
seen in the shape of ``dips'' in the tangential shear profile, and translate to
local mass excess in the lens convergence ($\kappa$) profile (see \citealt{Okabe2010};
see also \citealt{Zitrin2011c}, Figure 5). In the 2D WL analysis presented here,
these will simply be seen as 2D structures in the mass distribution, as can
be seen in Figure \ref{CC383WL}.

Note also that the detection here is based on the optical data,
meaning that the lensed cluster is clearly seen, and not by other
means which later need further verification.

In order to calculate, or put limits on the background cluster mass,
we first examine the total light by this cluster. The photometry of the 40
detected $z\sim0.9$ cluster members add up to a total B-band luminosity of
$0.84_{-0.19}^{+0.36} \times 10^{12}L_{\odot}$ (using the LRG templates from Ben\'itez
et al. 2009; errors include conversion from different bands), in roughly the $\sim2$
square-arcminutes occupied by this structure in the optical. After correcting for the
$14\pm3~ \%$ magnification effect, this yields a total source luminosity of
$0.74_{-0.17}^{+0.32} \times 10^{12} L_{\odot}$. Adopting
a typical $M/L_{B}$ ratio (e.g., \citealt{Zitrin2011a}), of 250$\pm50$ $M/L_{B}$,
this translates into a total mass of $\simeq 1.84^{+0.87}_{-0.56}
\times 10^{14} M_{\odot}$ for the lensed cluster, which constitutes a lower limit on
its mass, since we did not account for member galaxies lying below the detection threshold ($\simeq$26.5 AB mag in the $R_{c}$-band; $5\sigma$).

Next, we perform a 2D gravitational shear analysis of
Subaru multiband ($BVR_{c}I_{c}i'z'$) data (see Zitrin et
al. 2011b on A383) to constrain the mass distribution of the $z=0.9$ cluster.
The 2D-WL modelling procedure used in this work is
similar to that of \cite{Watanabe2011WL} and \cite{Okabe2011arXiv}. More details
will be presented in our forth-coming paper (Umetsu et al., in preparation).
From a photometric-redshift ($z_{\rm ph}$) selected sample of background
galaxies ($z_{\rm ph}>0.9$, $21<z'<26$\,AB mag), we construct a spin-2
reduced-shear field on a regular grid of $0.5\arcmin\times 0.5\arcmin$
independent pixels, covering a $20\arcmin \times 20\arcmin$ region
centred on A383.
We model the projected mass distribution around the $z=0.9$ background
cluster as the sum of two NFW halos, responsible for A383 and the
$z=0.9$ cluster, each parameterized with the halo virial mass ($M_{\rm
vir}$), concentration ($c_v$), and centroid position ($x_c,y_c$).
We use the Markov Chain Monte Carlo technique with
Metropolis-Hastings sampling to constrain the mass model with 8 parameters.

From a simultaneous 2-component fitting to the 2D shear data, we find
$M_{\rm vir}=1.51^{+1.45}_{-0.94}\times 10^{14}M_\odot$
($68.3\%$ confidence limits) for the $z=0.9$ cluster, where other model parameters
as well as the source redshift uncertainty are marginalized over.
With this model, the total projected mass within a (typical virial) radius of $R=1$ Mpc
is $M_{\rm 2D}\simeq 1.9\times 10^{14} M_\odot$. This result is in good agreement with
our optically-based estimation (although the latter constitutes a lower limit).

Additionally, we use SZ measurements to deduce the gas mass, from which we
estimate the total mass. Although the $z=0.9$ cluster is not detected in
the X-ray or SZ images, the SZ data do provide a constraint on the gas mass. The Bolocam SZ data were
processed using the procedure described in \cite{Sayers2011}; we discuss the relevant
aspects of this analysis below. In particular, the data are effectively high-pass
filtered in a complicated and slightly non-linear way in order to subtract noise due to
fluctuations in the opacity of the atmosphere (i.e., the transfer function
of the filtering depends weakly on the cluster profile). We fit an elliptical
generalized NFW profile (\citealt{Nagai2007pressureProfile,Arnaud2010pressureProfile}) to
A383, which provided a marginal fit quality ($\chi^2$/DOF = 1223/1117)
\footnote{Note that we simultaneously fit, and subtracted, a point source from the
NVSS catalogue at 02:48:22.09, -03:34:30.5, which we found, had a flux density of
12.8 mJy \citep{Condon1998NVSS}. This point source is approximately 5~arcmin from the
centres of both A383, and the $z=0.9$ cluster.}.

Using the transfer function computed for this model, we then deconvolved
the effects of noise filtering of our data to obtain an unbiased SZ image of
A383. Since the filtering is somewhat non-linear, our deconvolved image of the
$z=0.9$ cluster will be slightly biased because the transfer function used
for the deconvolution was determined from the profile of
A383. However, even in extreme cases the bias is $< 10$\% of the cluster
peak-height, which is negligible compared to our measurement noise for the
$z=0.9$ cluster.

In order to estimate the SZ signal from the $z=0.9$ cluster, we first
subtracted our best-fit model of A383 from our unbiased SZ image.
We then computed the integrated projected SZ signal within a 1~Mpc
aperture (2.14~arcmin) centred on the $z=0.9$ cluster, and found
$Y_{1Mpc} = 0.80 \times 10^{-11}$~ster, with a statistical error of
$\sigma_{Y,stat} = 1.06 \times 10^{-11}$~ster.
Restricting to physically allowed positive values of $Y_{1Mpc}$,
this results in a 95\% confidence level upper limit of
$Y_{1Mpc} < 2.66 \times 10^{-11}$~ster. Note that the model-subtracted
A383 SZ flux within our 1~Mpc aperture is $Y_{A383,model} = 1.89 \times 10^{-11}$~ster.

Using three published $Y-M_{gas}$ scaling relations based on projected $Y$
(\citealt{Sayers2011}, \citealt{Plagge2010YMrelation}, and
\citealt{Bonamente2008YMrelation}), we estimate the 95\% confidence level upper limit
for the
gas mass of the $z=0.9$ cluster within a spherical
region of radius of 1~Mpc to be 5.1, 6.6, and $2.7 \times 10^{13}$~M$_{\odot}$,
respectively. All three of these scaling relations were constrained largely
(or entirely) using clusters with much higher masses,
which is the likely cause of the large scatter in the
derived masses.
We adopt the gas mass limit
derived from the \cite{Sayers2011} scaling relation, since
it was calibrated using Bolocam data analysed identically to our
A383 data. We note that the spread in mass limits from the three scaling relations
provides an estimate of the uncertainties in the $Y/M$ scaling.

By assuming a typical gas mass fraction of 13\% (e.g., \citealt{Umetsu2009}),
we derive a 95\% confidence level upper limit
on the total mass of the $z=0.9$ cluster of $M_{SZ} < 3.9 \times 10^{14}$~M$_{\odot}$,
in agreement with our previous estimates based on the luminosity and our WL analysis.

For general completeness, we repeat a similar procedure also for the X-ray data, although only a rough upper mass limit can be expected. In order to estimate the X-ray surface brightness from the $z\sim0.9$ cluster, we consider X-ray observations retrieved from the Chandra archive and carried out with the ACIS--I CCD imaging spectrometer, namely observation ID 2320, with a total exposure time of approximately 20 ks, and use for our analysis the best-fit triaxial model of the IC gas of A383 obtained by \cite[][]{morandilimousin2011}. The IC gas model generates a theoretical surface brightness map which has been subtracted from the raw brightness image. We then integrate the differential counts within a 1 Mpc aperture centred on the $z = 0.9$ cluster, translated into an X-ray luminosity by assuming a metallicity of $Z=0.3$ solar value and the photoelectric absorption fixed to the Galactic value for the coordinates of the background cluster. This has been accomplished via fake {\it Chandra} spectra, where the emissivity model \citep[MEKAL model,][]{1992Kaastra, 1995ApJ...438L.115L} is folded through response curves (ARF and RMF) of the ACIS--I CCD imaging spectrometer in order to generate theoretical counts for the surface brightness. The luminosity value is obtained by rescaling the fake Chandra spectrum  in order to reproduce the observed number of differential counts.

Given that the X-ray luminosity mildly depends also on the value of the IC gas temperature of the background cluster, which is unknown a priori, we start by assuming a guessed value of the temperature and derive the corresponding luminosity. This luminosity is then translated via the $L-T$ relation \citep{morandi2007a} into a global temperature, which has been used iteratively in order to estimate the X-ray luminosity and then the same global temperature for the background cluster. This process is repeated until convergence of the temperature is achieved. The luminosity is finally converted into a total mass via standard X-ray scaling relations \citep[e.g.,][]{morandi2007a}, and errors on the physical parameters were calculated via Montecarlo randomisation. We derive a 95\% confidence level upper limit on the total mass of background cluster of $M_{\rm X}< 2.5\times 10^{14} M_{\odot}$, in agreement with our previous estimates.

\begin{figure}
  \centering
  \includegraphics[width = .40\textwidth]{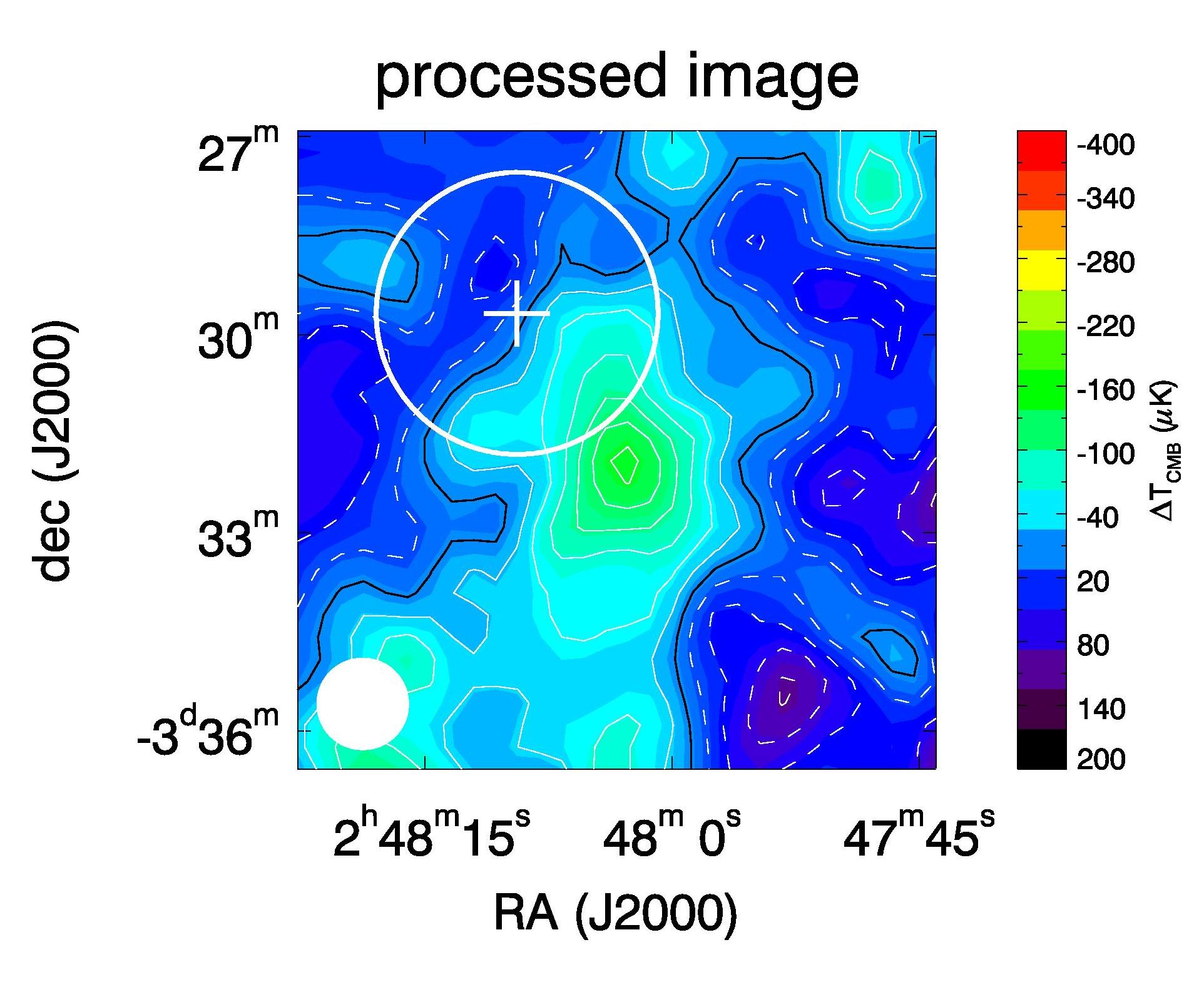}
  \includegraphics[width = .40\textwidth]{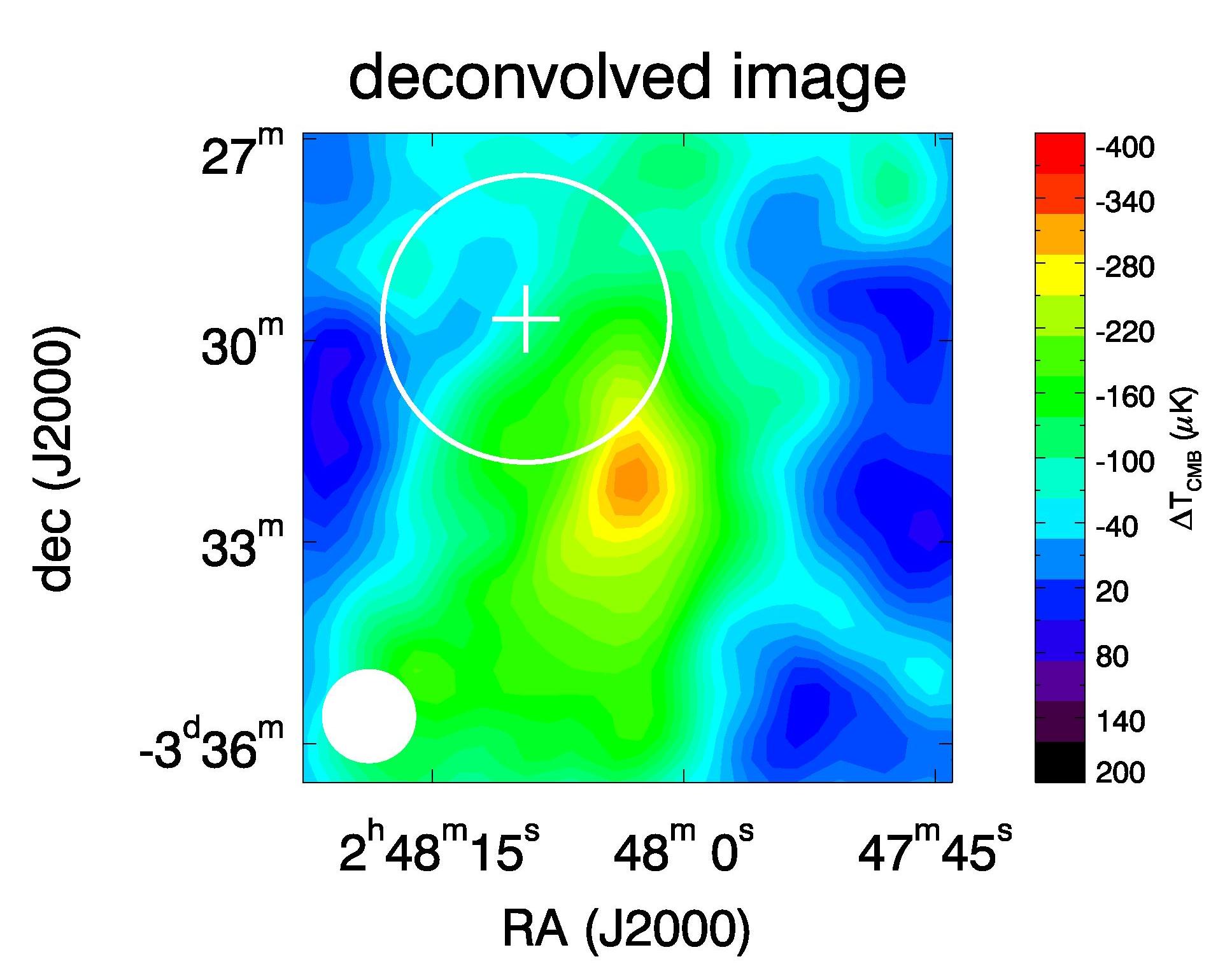}
  \includegraphics[width = .40\textwidth]{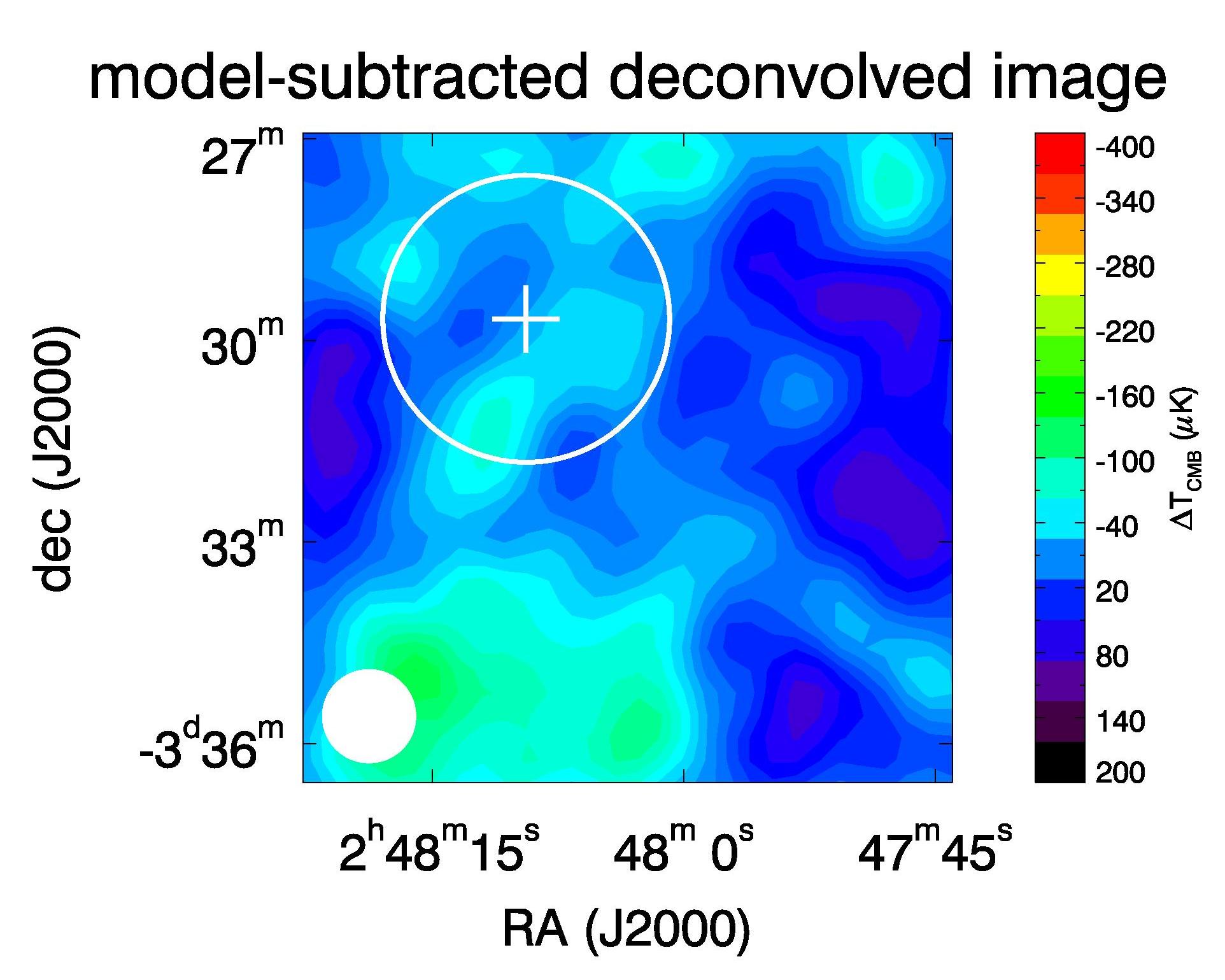}
  \caption{Bolocam SZ images of A383. From top to bottom the thumbnails show the
high-pass filtered (processed) image,
    the deconvolved image, and the model-subtracted deconvolved image.
    The solid white contours in the processed image represent
    S/N$ = -1, -2, -3,..$, and the dashed white contours
    represent S/N$ = +1, +2, +3,..$. We do not include S/N contours in the deconvolved
images
    due to the significant amount of large-angular-scale noise.
    The white plus sign denotes the centre of the $z=0.9$
    cluster, and the un-filled white circle denotes a
    radius of 1~Mpc at $z=0.9$ centred on the cluster.
    The solid white circle in the lower left represents
    the effective FWHM of the PSF in these beam-smoothed
    images. As is in the X-ray, the cluster lies below
the S/N threshold for detection, allowing only a 95\% confidence upper mass limit of $3.9 \times 10^{14}$~M$_{\odot}$, see
\S \ref{383}.}
\end{figure}

\section{Discussion and Summary}

We have calculated the predicted numbers of CCLs in the $\Lambda$CDM model for
different mass functions and cluster properties.
According to our rather conservative estimates, only few ($\sim 3$) CCLs are predicted over the full sky based on WMAP7 parameters,
using either a PS or a ST mass function, for clusters in the mass range
$1 \times 10^{13} - 1\times10^{16}$ $M_{\odot}$ (for both the lens and the source).
The number increases somewhat to $\sim10$ when taking into account also background
groups of galaxies (down to $5\times 10^{12}M_{\odot}$), and considering different mass
functions, but rises substantially to $\sim$ a few dozen when taking into account possible lensing triaxiality biases, and to hundreds when
considering also the weak lensing regime.

Two CCLs were claimed a decade ago - A2152, and less significantly, MS 1008-1224,
where background galaxies are obviously magnified or create a local over-density in
the image-plane. In addition, several ``double lens'' configurations were suggested or theoretically discussed before \cite[e.g.,][]{Crawford1986doubleLens,SeitzSchneider1994,Molinari1996doubleLens,Wang1997doubleLens,BertinLombardi2001CCL, Gavazzi2008doubleLens}. Comparison of the number of observed CCLs with theoretical predictions \cite[see also][]{Cooray1999CCL}
is clearly important and may add significant new insight on the evolution of the LSS
in $\Lambda$CDM. This could be quite useful in light of
claimed discrepancies, such as larger than predicted Einstein radii, and
high concentration or disparities in the abundance of giant arcs
\citep[e.g.,][]{Hennawi2007,BroadhurstBarkana2008,Broadhurst2008,SadehRephaeli2008,PuchweinHilbert2009,Horesh2010,Meneghetti2010a,Sereno2010,Zitrin2011a}.

While inspecting lensing measurements of a sample of clusters
we have noticed a lensed background clustering structure behind
A383 ($z=0.19$), in deep Subaru imaging. Photometric redshifts imply this
overdensity is at $z\sim0.9^{+0.2}_{-0.1}$, and is clearly seen redder in
an RGB colour image, and very faint in the B band. Our WL modelling of A383 implies a magnification of $14\pm3~\%$ of this background cluster.

Accounting for this correction,
the total B-band source luminosity is $0.74_{-0.17}^{+0.32} \times 10^{12}
L_{\odot}$ summed over all 40 members, which is translated into a lower-limit mass
estimate of $1.84^{+0.87}_{-0.56} \times 10^{14} M_{\odot}$, using a typical M/L ratio
of $M/L_{B}=250\pm50$. We have also analysed SZ and X-ray data, and independent WL measurements,
to obtain mass estimates of $M_{SZ} < 3.9 \times 10^{14}$~M$_{\odot}$, $M_{\rm X}< 2.5\times 10^{14} M_{\odot}$ (95\% CL upper
limits) and $M_{\rm vir, WL}=1.51^{+1.45}_{-0.94}\times 10^{14}M_\odot$ (or a projected mass
$M_{\rm 2D}(<1~Mpc)\simeq 1.9\times 10^{14} M_\odot$), respectively. These are commensurate
with our prior estimate based on the luminosity. These are also in agreement with the fact
that no excess emission is seen in the SZ or X-ray images of A383 at the location of the
background cluster (see Figure \ref{xray383}),
reflecting its probable low mass.

Deeper images in the different spectral regions would be of interest to examine
further the possibility that this cluster is a CCL, along with spectroscopic data
for measuring the exact background cluster redshift. In addition, future detections
of CCLs are important, as their overall number could probe the LSS parameters,
the cluster (and group) mass function, and the c-M relation.

\section*{acknowledgments}

AZ is grateful for the John Bahcall excellence prize which further encouraged this
work, and to Piero Rosati and members of the CLASH team for useful comments. Work at Tel Aviv University was partly supported by the US-IL Binational
Science foundation grant 2008452, and by a grant from the British Council. A.M. acknowledges support by Israel Science Foundation
grant 823/09. JS was partially supported by a NASA
Graduate Student Research Fellowship, a NASA Post-
doctoral Program fellowship, NSF/AST-0838261, and
NASA/NNX11AB07G; NC was partially supported by
NASA Graduate Student Research Fellowship. Bolocam observations and analysis were also supported by the Gordon and Betty Moore Foundation.
Part of this work is based on data collected at the Subaru Telescope, which is operated
by the National Astronomical Society of Japan. The \emph{Chandra} X-ray Observatory
Center is operated by the Smithsonian Astrophysical Observatory on behalf of NASA.
Bolocam was constructed and commissioned using funds from NSF/AST-9618798,
NSF/AST-0098737, NSF/AST-9980846, NSF/AST-0229008, and NSF/AST-0206158.


\bsp
\label{lastpage}

\end{document}